\documentclass[12pt,preprint]{aastex}
\newcommand{\skipthis}[1]{}
\def\nh3{$\rm{NH_3}$}
\def\NH3{$\rm{NH_3}$}

\def\kms-1{km~s$^{-1}$}
\def\h2o{$\rm{H_2O}$}
\def\h2{$\rm{H_2}$}

\newcommand{\lsim}{${\raisebox{-.9ex}{$\stackrel{\textstyle<}{\sim}$}}$ }

\begin{document}

\title {Warm molecular gas in galaxy-galaxy merger NGC6090}
\author{Junzhi Wang\altaffilmark{1,2,3}, Qizhou Zhang\altaffilmark{1}, Zhong Wang\altaffilmark{1}, Paul T. P. Ho\altaffilmark{1,4}, Giovanni G. Fazio\altaffilmark{1}, Yuefang Wu\altaffilmark{2}}
\altaffiltext{1}{Harvard Smithsonian Center for Astrophysics,
60 Garden Street,
Cambridge, Massachusetts 02138}
\altaffiltext{2}{Astronomy Department, Peking University, Beijing, 100871, China}
\altaffiltext{3}{e-mail: jwang@cfa.harvard.edu}
\altaffiltext{4}{Academia Sinica Institute of Astronomy and Astrophysics, P.O. Box 23-141, Taipei 106, Taiwan}
\begin{abstract}

We present observations of the CO 2-1 and 3-2 transitions
toward the merging galaxies of NGC6090
with the Submillimeter Array (SMA)\footnote[5]{The Submillimeter Array (SMA) is a joint
project between the Smithsonian Astrophysical Observatory and the
Academia Sinica Institute of Astronomy and Astrophysics, and is
funded by the Smithsonian Institution and the Academia Sinica.}. The
 high resolution CO data reveal three gas concentrations. 
The main component 
 is  peaking in the overlap region between the two galaxies, where the near-IR and radio continuum emission are weak. The CO 2-1 emission 
 from the face-on galaxy NGC6090E is somewhat stronger than that from the edge-on galaxy NGC6090W. 
 The CO 3-2 emission peaks in the overlap region, similar to the CO 2-1 emission. 
More than 50\% of
the CO 3-2 emission arises from the 2$''$ (1.2 kpc) area of the
overlap region.  There appears to be CO 3-2 
 emission toward the nuclear region and the north-west arm of NGC6090E, while no CO 3-2 emission
is detected toward NGC6090W. 
Unlike the CO gas, most of the radio continuum emission comes from NGC6090E.
The strong CO emission, together with the weak radio continuum emission, 
 suggests that star formation in the overlap region has not proceeded long enough to 
produce significant numbers of supernovae which would be detectable due to their radio continuum emission. 

\end{abstract}



\section{Introduction}

Galaxy-galaxy interactions and mergers may trigger starburst and
nuclear activity, and produce luminous/ultra-luminous infrared galaxies. 
The study of the  kinematics and
the distribution of molecular
 and atomic gas at high angular resolution can help us to understand how the gas responds to 
 merging and interacting processes. By studying different  transitions of CO emission, one
can determine the excitation conditions of the molecular gas.
Similarly, measurements of  millimeter/sub-millimeter continuum emission allow us to study the properties of
 dust. High-J CO transitions, such as CO 3-2 which traces the warm ($\sim$30K) and dense ($\sim$10$^4$cm$^{-3}$) molecular
gas, pinpoint sites of on-going star formation better than the low-J transitions. 
With these objectives  in mind, we mapped a gas rich merger  NGC6090  with
the partially completed SMA in the CO 3-2 and CO 2-1 transitions.

NGC6090 is a nearby IR-luminous galaxy-galaxy merger, at a distance of 122
 Mpc (using $H_0=75$ km s$^{-1}$ Mpc$^{-1}$,  Dinshaw et al. 1999), with
L$_{IR}(8-1000 \mu$$m)\simeq$$3\times10^{11} L_\odot$ (Dinshaw et al. 1999), 
M$_{H_2}\simeq$$3.0{\times}10^{10} M_\odot$ (Bryant \& Scoville 
1999), and M$_{HI}\simeq$$1.4{\times}10^{10}M_\odot$ (van Driel et al. 2001).
 It consists of two well separated nuclei seen at optical, near-IR, and
radio wavelengths.
 The projected separation is 5$''$.4 (3.2 kpc) measured at radio wavelengths (Dinshaw et
al.1999). 
There appears to be no evidence at the optical and radio wavelengths for an AGN, but
 there is clear evidence for starburst activity
 (Dinshaw et al.1999; Bryant \& Scoville 1999). 
CO 1-0 observations (Bryant \& Scoville 1999) show that the  dominant component of 
molecular gas peaks in the overlap region  between the two galaxies, where 
the near-IR emission and radio continuum  are weak (Dinshaw et al. 1999). 
Single dish CO 2-1 and CO 3-2 observations with the CSO (Glenn \& Hunter 2001) showed that the CO 2-1 and CO 3-2 
 are strong enough  to be  mapped  with the SMA. 
In order to understand the physical properties of the molecular gas, we
made high resolution images with the SMA in both the CO 2-1 and
CO 3-2 transitions. In Section 2, we describe instrumental parameters of the observations;
 in Sections 3 and 4, we present the results and compare 
this merger with other interacting systems; and in Section 5, we present
a brief summary.

\section{Observations}

Observations of NGC6090 were made using the partially completed SMA
during 2003. The CO 2-1 (observing frequency 223.929 GHz) data  were obtained on June 2, 2003, and August 4, 2003, with
4 and 6 antennas in the Array, respectively. 
The CO 3-2 (observing frequency 335.883 GHz) data were obtained on June 25, 2003, with 4 antennas. The typical system temperatures
 (SSB) are
220K and 300K for  CO 2-1  and CO 3-2 observations, respectively.
The projected
baselines ranged from 8m to 110 m.
 Two nearby quasars, 3C345 and 1637+574, were used to calibrate the 
 visibility amplitude and phase. During every 30-minute period, we integrated
on each of the calibrators for 5 minutes. The flux calibration was performed
with respect to Uranus, and was accurate to approximately 20\%.
The variations of gain across the 1 GHz spectral bands were monitored by observing  
Jupiter and/or Mars.  
The CO lines centered at 8600 km s$^{-1}$ (radio defined velocity) were placed in the 
 lower sideband. Because of the
large frequency shift, the strong CO lines in the Mars atmosphere
did not affect our bandpass calibration. We used a spectral resolution 
of 0.8 MHz, and smoothed the data to 21.4 \kms-1 for CO 2-1 when making the channel maps. 
 
The rms levels are 0.04 Jy beam$^{-1}$ and 0.06 Jy beam$^{-1}$
at  a velocity resolution of 21.4 km/s, for CO 2-1 (beamsize 2.4$''\times$1.5$''$ P.A. 62.2$^\circ$)
and CO 3-2 (beamsize 1.4$''\times$1.0$''$ P.A. 53$^\circ$), respectively.
 For a detailed description of the SMA, see Ho, Moran \& Lo(2004). 
 The SMA data were calibrated with the MIR package developed  at OVRO (Scoville et al. 1993).  The calibrated visibilities were
exported to MIRIAD for imaging. In order to compare the distribution
 of the integrated CO 3-2 and 2-1 emission, we convolved the CO 3-2 integrated intensity map 
to the same beam size as the  CO 2-1
. Continuum images were constructed from the image sideband.  

\section{Results}

Figure 1a shows the integrated intensity maps of the CO 3-2 and CO 2-1 lines overlaid
on the 1.49 GHz radio continuum map (Condon et al. 1990).
 Both CO 3-2
and CO 2-1 emission peak at the  overlap region between the two nuclei, similar to the CO 1-0 emission (Bryant \& Scoville 
1999).
The CO 2-1 emission appears to be  as extended as is the
CO 1-0 emission (Bryant \& Scoville 1999).  However, the CO 3-2 emission is more
concentrated toward the overlap region. The total CO 2-1
 flux  detected by the SMA  is 624 Jy km s$^{-1}$, similar to the value of the single 
dish flux from CSO (13.1 K km  s$^{-1}$, or 568 Jy km s$^{-1}$,
 Glenn \& Hunter 2001).  
The CO 3-2 emission is more compact and weaker than the CO 2-1 emission. More than 50$\%$ of the CO 3-2 flux is from the
overlap region, which only extends to an area of about 2$''$ (1.2 kpc). The total CO 3-2 flux from our map 
is 377 Jy km s$^{-1}$, about 30\% of the single dish flux from CSO (Glenn \& Hunter 2001).

We examined  the effect of missing short spacing in the SMA data. The CO 2-1 data from the SMA recovers nearly all the flux
measured by the CSO. The CO 3-2 data from the SMA have an inner hole in u,v plane of 9 k$\lambda$, $\sim$50\% larger than that in the CO 2-1 data.
We  flagged the CO 2-1 baseline shortward of 9k$\lambda$ thereby achieving the same 
 short spacing deficit as in the CO 3-2 data. The resulting CO 2-1 flux is 25\% less than
 the data without flagging.  We  expect that the CO 3-2 emitting region is less extended than the CO 2-1
 emitting region. 
 We therefore  estimate that the SMA images in CO 3-2 would  at most miss 25\% of the total flux.
 This is much less than the difference between the CSO flux and our SMA flux.
The discrepancy  between the CSO and the SMA results is mostly likely
 due to calibration  uncertainties at 345 GHz at the CSO (Hunter, private communication).

The brightness temperature ratio of the CO 2-1 from the SMA
 and the CO 1-0 from OVRO is 0.92. 
The ratio of the CO 3-2 and the CO 2-1 is about 0.27.  
And the  ratio of the CO 3-2 from the SMA and the CO 1-0 from OVRO is 0.25. The CO 3-2/1-0 ratio is much less than 
the mean value of 0.66 from the SCUBA survey of the local universe galaxies (Yao et al. 2003), which indicates  that the molecular gas in NGC6090 is cooler than  that in most of the galaxies in that sample.

The CO distribution is rather different from the
1.49 GHz continuum emission, which shows emission peaks associated with
the two galaxies (Condon et al. 1990). The 1.49 GHz 
emission is detected toward the peak of the
CO emission at a flux of 2.5 mJy/beam (beam size 1.5$''\times1.5''$), weaker than the peak flux of 6.5 mJy/beam toward NGC6090E,
or 3 mJy/beam toward NGC6090W. The two galaxies are also detected in the near-IR
with HST (Dinshaw et al. 1999). NGC6090E exhibits clear spiral arms and is likely
face-on, while  NGC6090W shows no spiral arm structures and is mostly
edge-on (P.A.$\simeq-10^\circ$). There is a lack of near-IR emission toward the peak of the CO emission
in the overlap region. Besides the main peak at the overlap region, there are 
 also two peaks in NGC6090E. One peak is at the nuclear region, the other is 
 at the north-western arm (about 2$''$ north and 1.5$''$ west of the nucleus) of NGC6090E. 
 There is also a secondary peak of CO 3-2 in the overlap region near the main peak, where the CO 2-1 emission is much weaker. 

Figure 1b shows the ratio of the integrated  CO 3-2 to 2-1  emission. We convolved
the CO 3-2 emission to the same beam size of the CO 2-1 line, and converted the
flux to units of temperature to remove the frequency dependence of the flux.
In nearly all the CO emitting region, the ratio 
of the CO 3-2 to 2-1 emission is 
less than 0.5. 
 There are three peaks in the
ratio map. The one toward the northwest of NGC6090E coincides
with a spiral arm seen in the near-IR (Dinshaw et al. 1999). Another peak is 
seen toward the nuclear region of NGC6090E. The third is at the CO 3-2 peak in the southern part of the overlap region.
Although there is no clear peak in the ratio map in the
overlap region, the ratio is still high. 

Figure 1c shows the mean velocity map (first moment) of the CO 2-1 emission.
There is a small velocity  gradient ($\sim10$km s$^{-1}$kpc$^{-1}$)
 toward the northwestern part of the face-on galaxy NGC6090E. On the other hand, the velocity 
gradient is larger toward the southeastern part of NGC6090E, and also
toward the overlap region ($\sim30$km s$^{-1}$kpc$^{-1}$) . These results are similar to
those of the CO 1-0 emission (Bryant \& Scoville 1999).   The velocity and intensity distribution of the CO 2-1 emission 
in the overlap region indicate that it is possibly a rotating gas disk (P.A.$\simeq70^\circ$).
 Because NGC6090E is a face-on galaxy
 and NGC6090W is an edge-on galaxy with P.A.$\simeq-10^\circ$, the gas component does not appear to  belong to either NGC6090E or NGC6090W based on the kinematics.  

Figure 1d shows the line width (second velocity moment)  map of the CO 2-1 emission.
The line width peaks at the spiral arm of NGC6090E where there is a CO 3-2 peak as well. 
 Line width also peaks toward the overlap 
region as it has been seen in CO 1-0 (Bryant \& Scoville 1999). It is interesting to note that the line widths toward the overlap region
are larger than most parts toward both galaxies.   
In particular, NGC6090W is nearly
edge-on.  

Figure 2 shows the CO 2-1 channel maps. The reference velocity 
 is 8600 km s$^{-1}$. Most of the CO 2-1 emission is in the velocity range between -74.9 km s$^{-1}$ and 74.9 km s$^{-1}$. 
  CO 2-1 emission can be seen in almost all the 8 channels from  -74.9 km s$^{-1}$ to 74.9 km s$^{-1}$ in the overlap region. 
 The concentration of the face-on galaxy NGC6090E can be seen in 3 channels, from -10.7 km s$^{-1}$ to 32.1  km s$^{-1}$. The 
concentration of the edge-on galaxy NGC6090W is distributed over several channels from -96.3  km s$^{-1}$ to 32.1  km s$^{-1}$, and
 mixes with the main component.

No continuum was detected in either 230 GHz or 345 GHz at an rms of 4 mJy 
and  8mJy, respectively. The total 1.3mm (230 GHz) continuum was 20 mJy as detected by CSO (Glenn \& Hunter 2001). The 230 GHz continuum must 
 be more extended than two beams of our 230 GHz map. Otherwise, we should 
have detected it with the SMA.

\section {Discussion}

 The CO 2-1 emission shows gas concentrations toward the  overlap region 
 and the two galaxies. The kinematics of the molecular 
 gas seen in the CO 2-1 mean velocity map (see Figure 1c) indicates that  most of the gas 
 has settled in the overlap region and probably it is  a newly formed gas disk. 
There exists CO 2-1 emission toward NGC6090E with very low velocity gradient ($\sim10$km s$^{-1}$kpc$^{-1}$).
 In addition, CO 2-1 emission in NGC6090W can be found in the channel maps in Figure 2.
 The near-IR overlap bridge is 2$''$ (1.2kpc) west of the CO peak (Dinshaw et al. 1999).  
In interacting systems, the merger of gas is more efficient than that of stars.
Furthermore, the gas may evolve differently from stars. 
For example, although the two galaxies are well separated (3.4 kpc), it seems that most of the molecular gas in NGC6090
 has condensed in the overlap region.
 The CO 3-2 map shows that most of the warm and dense gas is segregated toward the overlap region, which suggests 
 that star forming activity in this merger is located at the overlap region rather than  the circum-nuclear regions.

   The star formation efficiency, which can be defined as the ratio of the IR luminosity to the molecular gas mass, 
  is only about 10$L_{\odot}/M_{\odot}$ in NGC6090. This value is similar to those of  the nearby starburst galaxies, but 
 lower than those of the luminous galaxies (Sanders \& Mirabel 1996). Low star 
formation efficiency ($\lsim$ 10$L_{\odot}/M_{\odot}$) is common in gas-rich, less advanced mergers, 
 such as VV114 (Iono et al. 2004), NGC4038/9 (Gao et al. 2001), 
 NGC6670 (Wang et al. 2001), Arp320 (Lo et al. 1997), and UGC12914/15 (Gao et al. 2003). 
In NGC6670, a system at an earlier merging stage than NGC6090, the
gas in the overlap region is mostly atomic rather than molecular 
(Wang et al. 2001).
 If NGC6090 and NGC6670 are similar in terms of the merging process, but at a different merging stage,
  some of the molecular gas in the overlap region
  might originate from atomic gas of the two galaxies in NGC6090.

The distribution of the CO 2-1 emission in NGC6090 is similar to that in
 VV114: The CO 2-1 emission resembles the CO 1-0 and peaks toward the overlap region (Iono et al. 2004, Yun et al. 1994). 
However, the CO 3-2 data in NGC6090 shows a different warm gas 
 distribution from that of VV114. Unlike NGC6090, the CO 3-2 emission in VV114 does 
not peak toward the overlap
  region, but arises from one of the two galaxies (Iono et al. 2004). 
The CO emission in NGC6090 is more similar to the Antennae galaxies (NGC4038/9). Although the optical morphology of NGC6090 and
 the Antennae galaxies  are different, the CO 1-0, 2-1 and 3-2 emissions all peak at the overlap
 region (Wilson et al. 2000, Gao et al. 2001, Zhu et al. 2003).  
 Even though most of the  CO 3-2 flux in NGC6090 comes from the overlap region, the weak radio continuum indicates 
that the history of star formation may not be very long. Otherwise, supernovae 
after the starbursts would have 
 produced  comparable radio continuum emission as in the two nuclei. 
Because of the presence of warm and dense gas,
 we expect higher infrared luminosities and  radio continuum 
 in the overlap region caused by on-going starbursts to emerge in the near future (possibly within $10^6$yrs, the main-sequence lifetime of massive stars). The
CO 3-2 emission also indicates the presence of star forming activity  
in the nuclear  region of NGC6090E.  The weak radio continuum and weak
 CO 3-2 emission in NGC6090W suggest that the star formation activity there is much less than that in NGC6090E.

\section {Summary}

Observations of the CO 2-1 and 3-2 transitions toward NGC6090 allow us to 
isolate dense and warm gas in the overlap region of the two galaxies.
The CO 2-1 velocity width  shows a peak toward the overlap 
 region and another peak toward the spiral arm of NGC6090E. The CO 3-2 emission
is found in the  nuclear  region of NGC6090E and the overlap region which
implies active star formation there.
The presence of warm and dense molecular gas,
together with  weak near-IR and radio continuum indicates that the epoch
of active star formation  in the overlap region must be fairly recent. 
No significant CO 3-2 
 emission can be seen toward the edge-on galaxy NGC6090W. 
The warm and dense molecular gas traced by  the CO 3-2 in NGC6090 reveals a
 different structure from the total molecular gas traced by  CO 2-1 and CO 1-0. 
This study illustrates the usefulness of high angular resolution imaging at sub-millimeter wavelengths of the CO 3-2 line 
to study the physical
properties of the molecular gas in merging systems.

 We thank Todd R. Hunter for sharing the CSO CO 3-2 and CO 2-1 spectra. We also thank Yu Gao, Jim Moran, Chunhua Qi and  Di Li for helpful discussions and comments. We thank the anonymous referee for his/her great help. Junzhi Wang and Yuefang Wu acknowledge 
 the support of NSFC grants 10128306 and 10133020.
 We also thank all the SMA staff of SAO and ASIAA.

\clearpage

\centerline {\bf Figure Captions}

{\bf Figure 1 (a).} Moment zero map of the CO 2-1 (red contours) and CO 3-2 (green contours) obtained
with the SMA overlaid on the 1.49 GHz continuum map from the VLA (Condon et al. 1990). 
Contours of both CO 2-1 and CO 3-2 are  in steps of 10 Jy beam$^{-1}$ km s$^{-1}$.
Beam sizes,  at the bottom left and right of the panel, are  for CO 2-1 (uniform-weighted)
data and CO 3-2 data (natural-weighted), respectively.  
The two star symbols mark the 1.49 GHz continuum peaks: $\alpha=$16$^h$11$^m$40$^s$.889, $\delta=$52$^\circ$27$'$26$''$.51 (J2000); and $\alpha=$16$^h$11$^m$40$^s$.418   $\delta=$52$^\circ$27$'$23$''$.86 (J2000).
 {\bf (b).} Ratios in brightness temperature of the integrated CO 3-2 and 2-1 emission (grey scales)
in comparison with the integrated CO 2-1 emission in contours. The CO 3-2
data are convolved to the beam size of CO 2-1.
 The gray scales are from 0 to 0.5. 
 {\bf (c).} The intensity weighted mean velocity (first moment) map of the CO 2-1 emission.
The gray scales are from -30 km s$^{-1}$ to 100 km s$^{-1}$ (reference velocity 8600 km s$^{-1}$).
 The contours are -30 (dot), -20 (dot-dash), -10 (dash), 0, 10, 20, 30, 40, 50 km s$^{-1}$.  
 {\bf (d).} The line-width (second velocity moment) map of the
CO 2-1 emission (grey scales)
in comparison with the integrated CO 2-1 emission in contours. Gray scales are plotted from 10 to 50  km s$^{-1}$. The contours 
for the CO 2-1 are the same as those in (a).

{\bf Figure 2.} Channel maps (reference velocity 8600 km s$^{-1}$) of the CO 2-1 line,
smoothed to 21.4 \kms-1 velocity resolution. Beam size (2.4$''\times$1.5$''$ P.A. 62.2$^\circ$) is marked by the shaded
ellipse at the lower left corner of the first panel. The 
contours are plotted in steps of 0.12Jy beam$^{-1}$ (3$\sigma$). 
 The two star symbols mark the 1.49 GHz continuum peaks.

\clearpage

\begin{figure}[h]
\begin{center}

\includegraphics{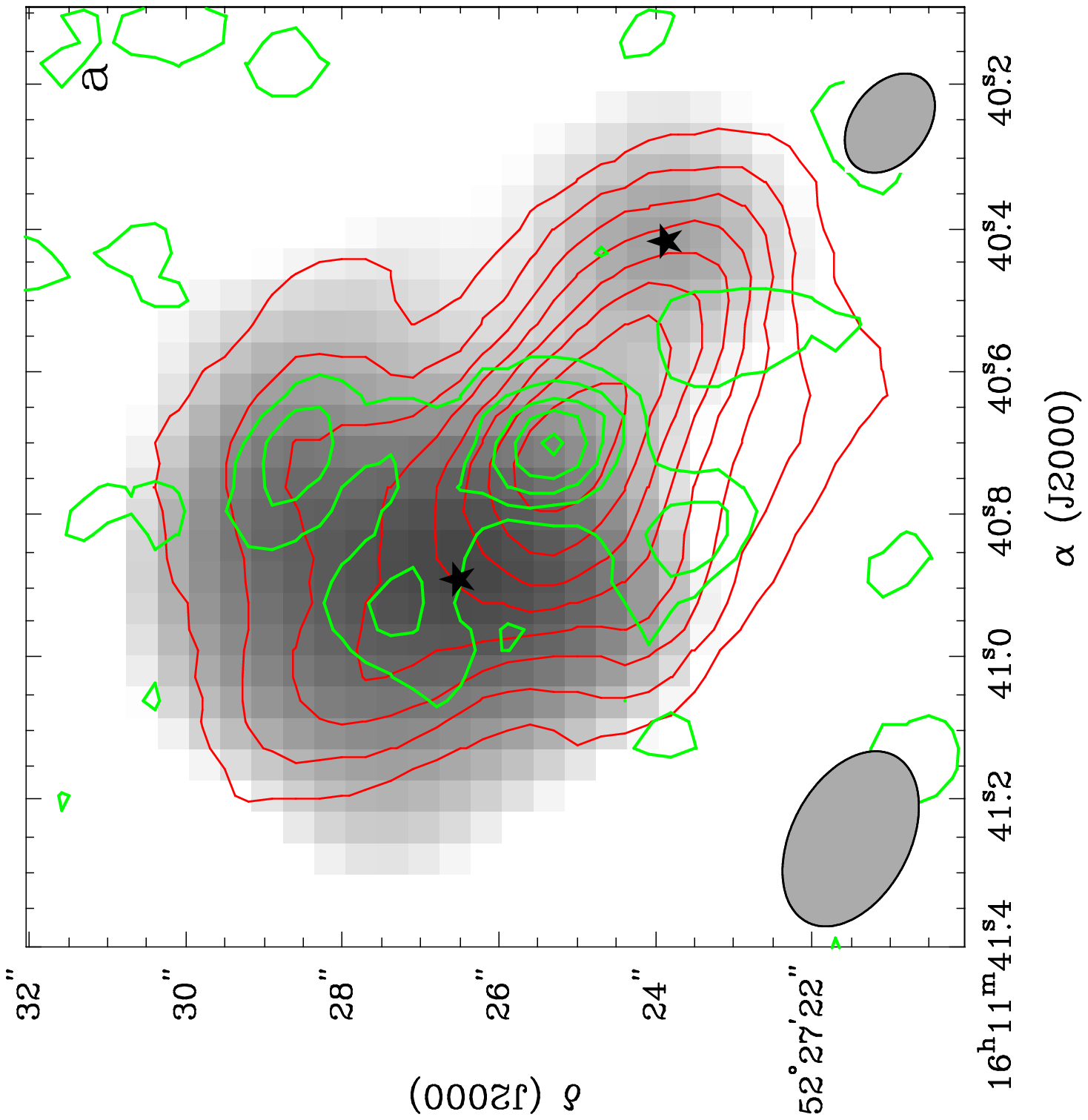}

\includegraphics{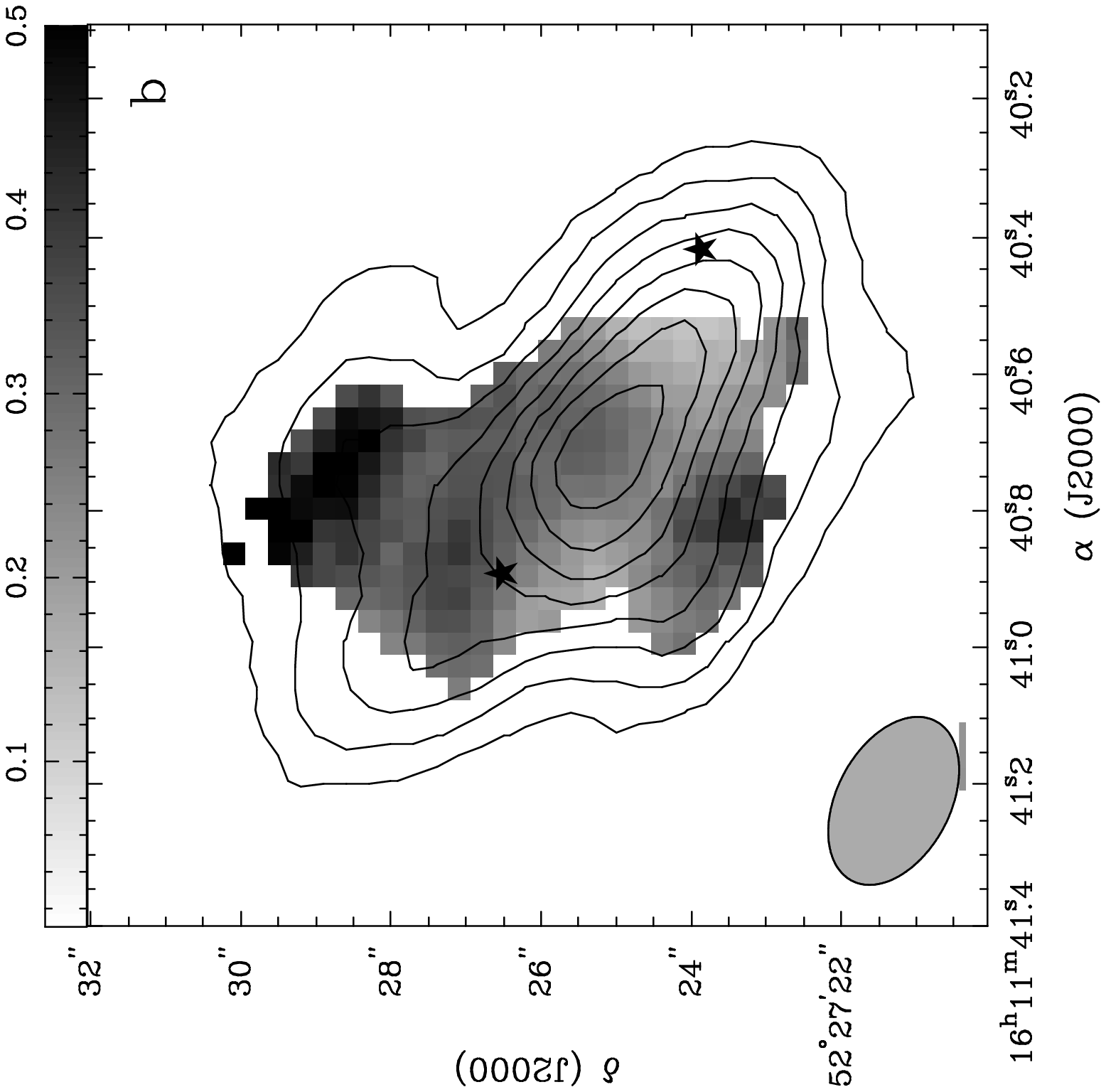}
\includegraphics{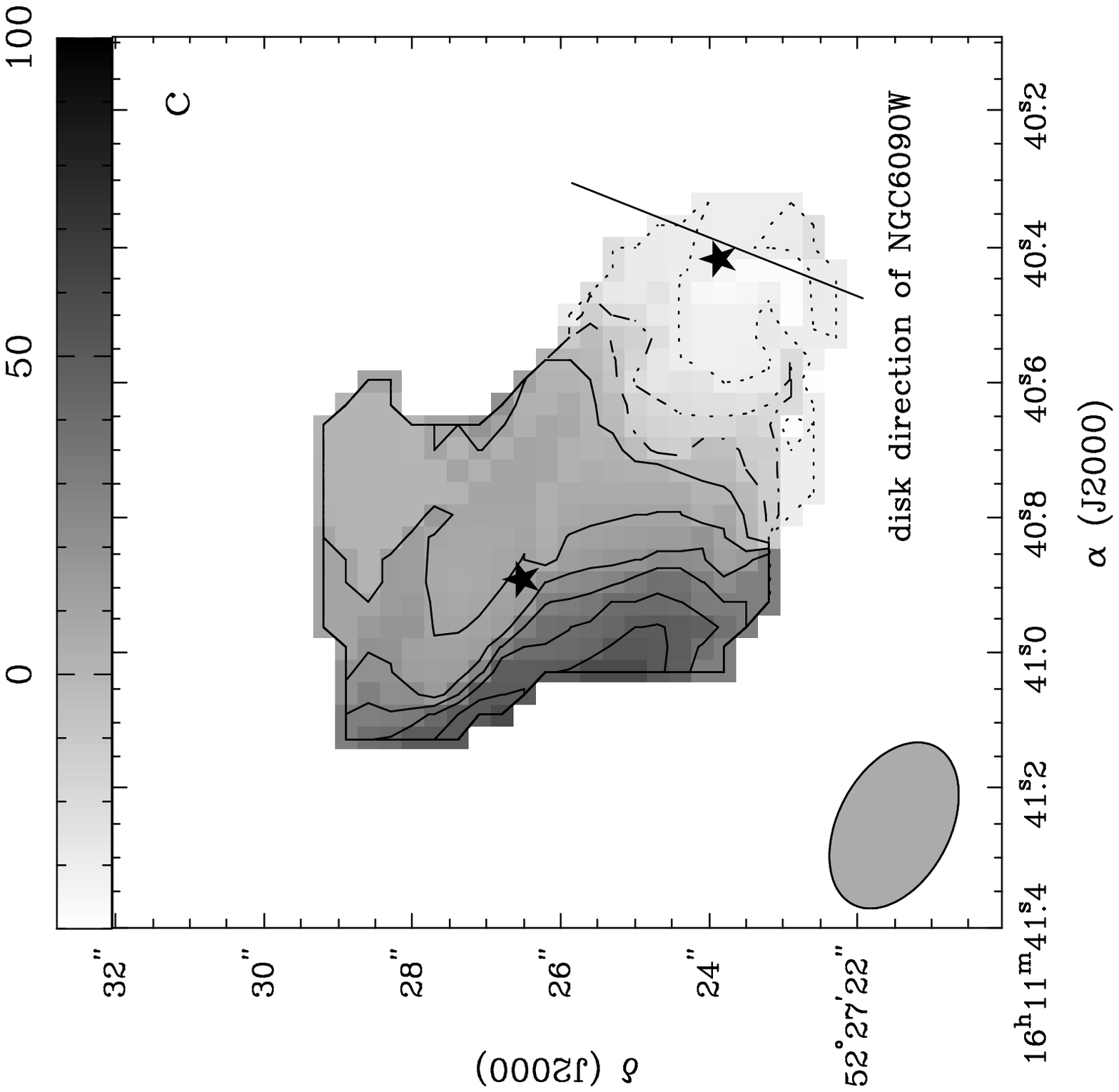}
\includegraphics{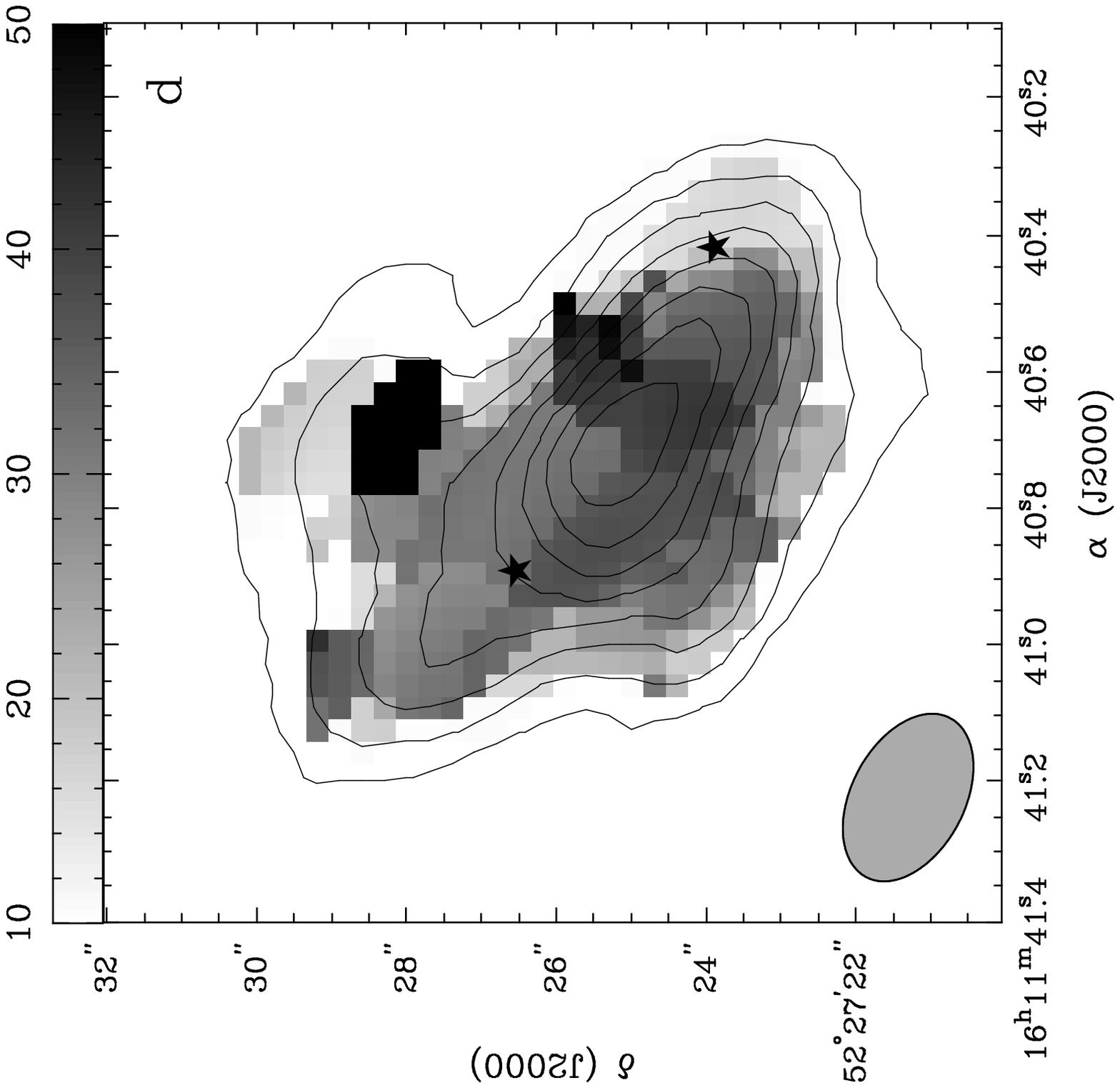}

\end{center}

\caption{}
\end{figure}

\clearpage

\begin{figure}[h]
\begin{center}
 
\includegraphics{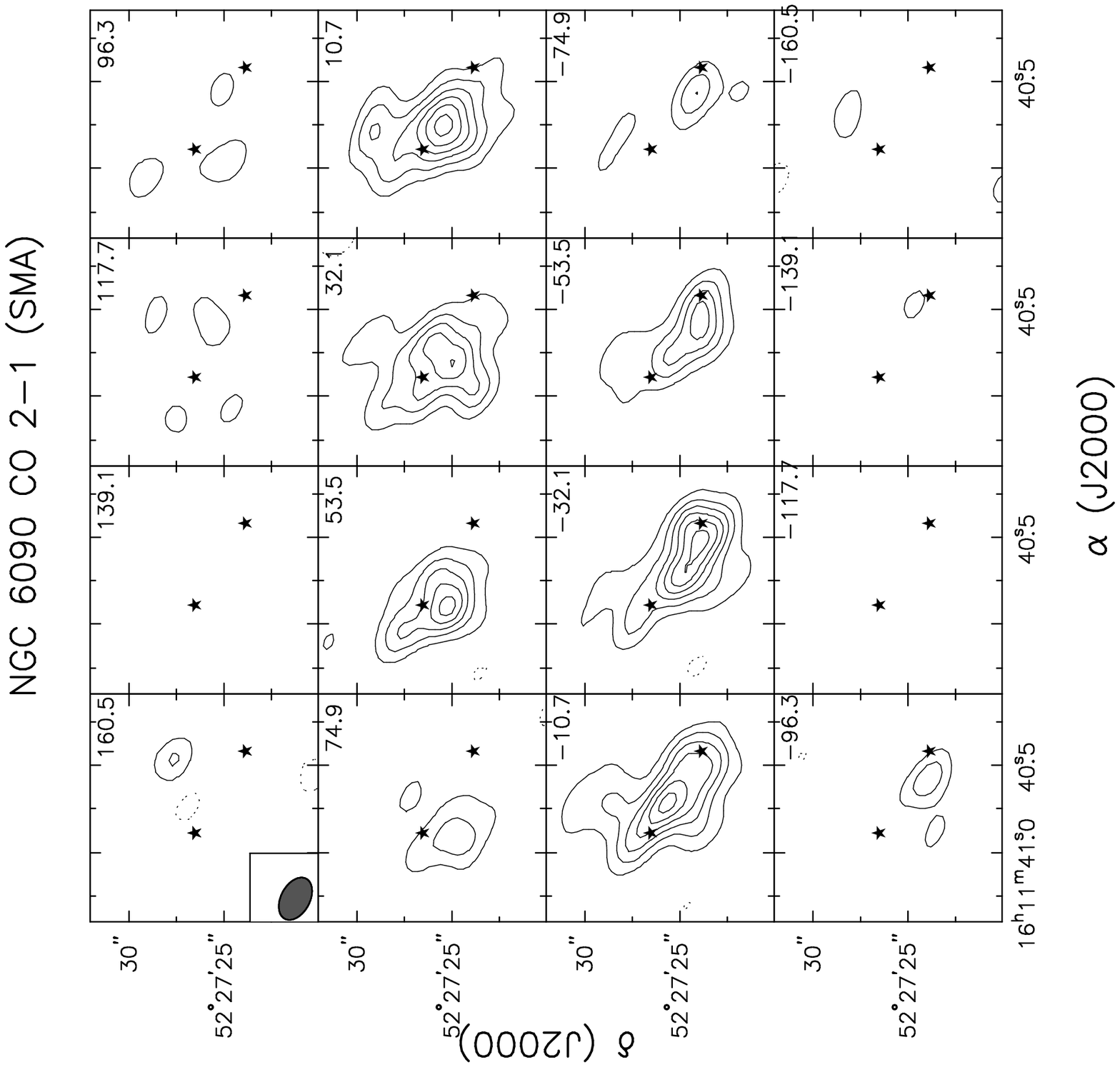}

\end{center}

\caption{ }
\end{figure}

\end{document}